\documentclass{aa}
\input epsf
\newcommand{\nha}{$N_{\rm H}$}
\newcommand{\nh}{$N_{\rm H}$~}

\begin{document}

   \thesaurus{03     
              (11.01.2;  
               11.09.4;  
               11.19.1;  
               11.19.7;  
              )} 
   \title{Nuclear obscuration and scattering in Seyfert 2 galaxies}

   \author{Qiusheng Gu \inst{1,2}
           \and
           Roberto Maiolino \inst{3}
           \and
           Deborah Dultzin-Hacyan \inst{2}
          }

   \offprints{Q. Gu, email: qsgu@nju.edu.cn}

   \institute{ 
            Department of Astronomy, Nanjing University,
            Nanjing 210093, P.R. China
         \and
            Instituto de Astronom\'{\i}a,
            Universidad Nacional Aut\'onoma de M\'exico, 
            Apdo Postal 70-264, Mexico D.F. 04510,
            M\'exico 
         \and
            Osservatorio Astrofisico di Arcetri, Largo Enrico Fermi 5, 
            I $-$ 5015 Firenze, Italy
     }

   \date{Received --- ; accepted --- }

   \titlerunning{Obscuration and Scattering in Seyfert 2s}

   \authorrunning{Gu, Maiolino \& Dultzin-Hacyan}

   \maketitle

   \begin{abstract}

 We study the relation between gaseous absorbing column density (\nha),
 infrared colors and detectability of the broad lines in a large sample
 of Seyfert 2 galaxies(Sy2s). We confirm that Sy2s without polarized 
 broad lines tend to have cooler 60$\mu$m/25$\mu$m colors; this correlation 
 was previously ascribed to the
 effect of obscuration towards the nuclear region.
 We find some evidence that Sy2s without polarized broad lines
 have larger absorbing column density (\nha) and that a fraction 
 of them are characterized by dust lanes crossing their nuclei. 
 However, we find that the IR colors do not
 correlate with \nha, in disagreement with the obscuration scenario.
 Also, Sy2s without polarized broad lines follow the same radio-FIR relation as
 normal and starburst galaxies, at variance with Sy2s with polarized broad
 lines. These results indicate that the lack of broad lines in the
 polarized spectrum of Sy2s is mostly due to the contribution/dilution
 from the host galaxy or from
 a circumnuclear starburst, though at a lower extent the
 obscuration toward the nuclear region also
 plays a role.

   \end{abstract}

\section{Introduction}

According to the standard unification model, Seyfert 1 and 2 galaxies
(Sy1s and Sy2s hereafter) are intrinsically the same objects and 
the absence of broad lines in Sy2s is ascribed to the
obscuration by a pc-scale dusty torus oriented along the line of sight
(see the reviews by  Antonucci \cite{antonucci_rev} and 
V\'eron-Cetty \& V\'eron  \cite{veron}).
The observational evidence for this model
includes the detection of polarized broad 
emission lines in some Seyfert 2 galaxies (Antonucci \& Miller
\cite{antonucci}; Tran \cite{tran}; Moran et al. \cite{moran}), the detection
of broad lines in the infrared spectrum of some Sy2s (Ruiz et al.
\cite{ruiz}; Veilleux et al. \cite{veilleux}; Rix et al.  \cite{rix}) and 
the detection of a prominent
photoelectric cutoff in the X-ray spectra of Sy2s indicating the presence
of large columns of gas along the line of sight
(Koyama et al. \cite{koyama}; Awaki et al. \cite{awaki};
Maiolino et al. \cite{maiolino98}; Risaliti et al. \cite{risaliti}).

The spectropolarimetric observations of different
samples of Seyfert 2 galaxies indicate that {\it only} about 40 \% of
Sy2s show broad lines in their polarized spectra
(eg. Heisler et al. \cite{heisler}), although such
surveys are probably biased since pre-selection was done according to the
broad-band polarization.

According to the suggestion of Heisler et al. (\cite{heisler}),
the detectability of a hidden BLR through spetropolarimetry in Sy2s
is related to the inclination of the torus which, in turn, is related 
to the 60 $\mu$m to 25 $\mu$m flux ratio, $s_{60\mu m}/s_{25\mu m}$.  
More specifically, in those Sy2s showing polarized broad lines 
(PBL) the torus should be oriented more
face-on so that the scattering medium is less obscured by the torus itself.
This model would also explain the correlation between IR colors
($s_{60\mu m}/s_{25\mu m}$) and detectability of the PBL. In particular,
when the torus is observed close to pole-on, the PBL should be more
easy to detect and the IR color should be hotter since we are observing the
hotter dust emitting region, in agreement with what is observed.
 
However, more recently Alexander (\cite{alexander}) compared the absorbing 
column densities inferred from the hard X-rays with the detectability of the PBL
and found no correlation. This result is in contrast with Heisler's et al.
(\cite{heisler}) model, which would predict a higher absorbing column
for Sy2s without PBL. Alexander (\cite{alexander}) suggests that the relation between
detectability of the PBL and IR colors is indirect: the contribution from
the host galaxy would both make the IR color
cooler and would also dilute the nuclear optical spectrum making more difficult
the detection of scattered polarized light.

The simple model of the obscuring torus has been subject to various
modifications. In particular, while the pc-scale torus is probably
responsible for the huge absorbing columns (\nh $> 10^{24}$ cm$^{-2}$)
observed in several Sy2s, observational evidence was also found
for a larger scale ($\sim$ 100 pc) obscuring medium
with lower absorbing column density (\nh $\sim$ a few times $10^{22}$ cm$^{-2}$,
Granato et al. \cite{granato}; Matt \cite{matt}; Maiolino \cite{maiolino} 
and references therein).

In this paper we expand the work done by Alexander (\cite{alexander}) by 
enlarging
the sample of Sy2s for which information on both the absorbing \nh and on
the detection of PBL is available. We also seek additional constraints on
the nature of the circumnuclear scattering and absorbing medium by
comparing the mid- and far-IR colors with \nh and the
detectability of PBL with the nuclear morphology and with the radio power.

The paper is organized as follows. In Section 2 we present our sample of
Seyfert 2 galaxies with spectropolarimetric observations, the results are
given in Section 3. We discuss our results and their implications in
Section 4 and summarize our conclusions in Section 5.


\section{The Sample}

 We collected all Seyfert 2 galaxies from the recent literature (from
 1985 to 2000), for which both spectropolarimetric
 data and an estimate of \nh from the X-rays are available.
 Within this sample, we got 22 Seyfert 2 galaxies with PBL, and 18 Sy2s 
 without detection of PBL, which are presented in Table 1 
 and 2, respectively.

In Table 1 and 2, we report the following information:
galaxy name (column 1); column density (\nha) taken 
from Bassani et al. (\cite{bassani}); Risaliti et al. (\cite{risaliti});
and Alexander (\cite{alexander}) (column 2); 
the IRAS colors s$_{25\mu m}/s_{12\mu m}$ and s$_{60\mu m}/s_{25\mu m}$ 
in columns 3 and 4, 
where the IRAS fluxes are taken from Moshir et al. (\cite{moshir}); 
and the flux between 42.5 and 122.5 $\mu$m, FIR, where 
FIR $\rm = 1.26\times 10^{-14} (2.58\times s_{60\mu m} + s_{100\mu m})$,
in column 5; the 1.49 GHz radio emission from the NRAO/VLA Sky Survey (NVSS)
(Condon et al. \cite{condon}) in column 6; 
and the corresponding reference for PBL in column 7.

\begin{table*}
\caption{Basic Data for Seyfert 2 Galaxies with PBL}
\begin{tabular}{lcrccrrl}
\hline
\hline
Name & &$N_{\rm H}$$^{\rm a}$&s$_{25\mu m}/s_{12\mu m}$&s$_{60\mu m}/s_{25\mu m}$&FIR$^{\rm b}$&F$_{\rm 1.49 GHz}$$^{\rm c}$&Ref\\
\hline
  Circinus      &    & 43000$^{+19000}_{ -11000}$ & 3.640 &  3.634  & 12.065 &        & ${\rm h}$  \\
  ESO 434$-$G40 &    &   162$^{+  23}_{ -   21}$  &       &         &        &  14.6  & ${\rm h}$  \\
  F05189$-$2524 &    &   490$^{+   10}_{ -   16}$ & 4.632 &  3.960  &  0.600 &   29.1 & ${\rm h}$  \\
  F09104$+$4109 &    &    24$^{+    6}_{ -    6}$ & 1.520 &  1.395  &  0.030 &        & ${\rm j}$  \\
  F13197$-$1627 &    &  7943                      & 3.135 &  2.047  &  0.259 &  275.3 & ${\rm d}$  \\
  F20460$+$1925 &    &   250$^{+  34}_{ -   32}$  & 1.385 &  1.685  &  0.042 &  18.9  & ${\rm h}$  \\
  F23060$+$0505 &    &   840$^{+ 190}_{ -  250}$  & 1.438 &  2.500  &  0.051 &   6.8  & ${\rm h}$  \\
  IC 3639     & $>$&100000                        & 3.538 &  3.130  &  0.374 &        & ${\rm i}$  \\
  IC 5063     &    &  2400$^{+  200}_{ -  200}$   & 3.310 &  1.557  &  0.250 &        & ${\rm k}$  \\
  Mark    3    &    & 11000$^{+ 1500}_{ - 2500}$  & 4.057 &  1.377  &  0.171 & 1100.9 & ${\rm e}$ \\
  Mark  348    &    &  1060$^{+  310}_{ -  260}$  & 2.484 &  1.870  &  0.070 &  292.7 & ${\rm e}$ \\
  Mark  463E   &    &  1600$^{+  800}_{ -  800}$  & 2.825 &  1.354  &  0.094 &  381.0 & ${\rm e}$ \\
  Mark  477    & $>$& 10000                       & 2.160 &  2.500  &  0.067 &   60.8 & ${\rm f}$ \\
  Mark 1210    & $>$& 10000                       & 3.800 &  0.880  &  0.079 &  114.9 & ${\rm f}$ \\
  NGC 1068    & $>$&100000                        & 2.267 &  2.140  &  9.070 & 4849.0 & ${\rm g}$ \\
  NGC 2110    &    &   289$^{+   21}_{ -   29}$   & 2.378 &  5.068  &  0.224 &  299.4 & ${\rm h}$  \\
  NGC 2273    & $>$&100000                        & 2.957 &  4.654  &  0.335 &   63.4 & ${\rm i}$  \\
  NGC 3081    &    &  6600$^{+ 1800}_{ - 1600}$   &       &         &        &    5.7 & ${\rm o}$  \\
  NGC 4388    &    &  4200$^{+  600} _{- 1000}$   & 3.550 &  3.070  &  0.578 &  120.4 & ${\rm h}$  \\
  NGC 4507    &    &  2920$^{+  230} _{-  230}$   & 3.065 &  3.248  &  0.219 &  67.4  & ${\rm o}$  \\
  NGC 5506    &    &   340$^{+   26} _{-   12}$   & 2.800 &  2.420  &  0.409 &  339.4 & ${\rm h}$  \\
  NGC 7674    & $>$&100000                        & 2.667 &  2.901  &  0.288 &  221.4 & ${\rm e}$ \\
\hline
\end{tabular}
\end{table*}

\begin{table*}
\caption{Basic Data for Seyfert 2 Galaxies without PBL}
\begin{tabular}{lcrccrrl}
\hline
\hline
Name & &$N_{\rm H}$$^{\rm a}$&s$_{25\mu m}/s_{12\mu m}$&s$_{60\mu m}/s_{25\mu m}$&FIR$^{\rm b}$&F$_{\rm 1.49 GHz}$$^{\rm c}$&Ref\\
\hline
  F19254$-$7245 &    &  1995$^{\rm d}$            & 5.462 &  3.690  &  0.249 &        & ${\rm l}$   \\
  Mark 1066    & $>$& 10000                       & 4.620 &  4.524  &  0.505 &        & ${\rm o}$   \\
  NGC   34    & $>$&  1000$^{\rm d}$              & 5.667 &  7.155  &  0.779 &   67.5 & ${\rm l}$   \\
  NGC 1143    & $>$&   100$^{\rm d}$              & 2.692 &  7.600  &  0.319 &        & ${\rm l}$   \\
  NGC 1386    & $>$&100000                        & 2.880 &  4.090  &  0.316 &  37.8  & ${\rm o}$   \\
  NGC 1667    & $>$& 10000                        & 1.763 &  8.731  &  0.375 &  77.3  & ${\rm o}$   \\
  NGC 3281    &    &  7980$^{+ 1900} _{- 1500}$   & 2.909 &  2.641  &  0.317 &  80.9  & ${\rm o}$   \\
  NGC 3393    & $>$&100000                        & 2.840 &  3.352  &  0.127 &   81.5 & ${\rm m}$   \\
  NGC 4941    &    &  4500$^{+ 2500} _{- 1400}$   & 2.280 &  2.351  &  0.095 &  20.3  & ${\rm o}$   \\
  NGC 5128    &    &  2250$^{+ 1250} _{- 1250}$   & 1.346 & 11.441  &  9.829 &        & ${\rm n}$   \\
  NGC 5135    & $>$& 10000                        & 3.701 &  6.524  &  0.914 &   1.5  & ${\rm l}$   \\
  NGC 5347    & $>$& 10000                        & 3.172 &  1.565  &  0.081 &        & ${\rm o}$   \\
  NGC 5643    & $>$&100000                        & 3.895 &  5.585  &  1.165 &        & ${\rm o}$   \\
  NGC 7130    & $>$& 10000                        & 3.397 &  7.790  &  0.873 &  190.6 & ${\rm l}$   \\
  NGC 7172    &    &   861$^{+   79} _{-   33}$   & 1.696 &  7.641  &  0.356 &   37.6 & ${\rm l}$   \\
  NGC 7496    &    &   501$^{\rm d}$              & 5.630 &  5.625  &  0.471 &        & ${\rm l}$   \\
  NGC 7582    &    &  1240$^{+   60} _{-   80}$   & 4.689 &  7.585  &  2.477 &        & ${\rm l}$   \\
  NGC 7590    & $<$&   9.2                        & 1.615 &  8.798  &  0.467 &        & ${\rm l}$   \\
\hline
\end{tabular}

\noindent
$^{\rm a}$ Absorbing column density in unit of 10$^{20}$ cm$^{-2}$;
$^{\rm b}$ In unit of 10$^{-12}$ W m$^{-2}$; 
$^{\rm c}$ In unit of mJy; 
$^{\rm d}$ Alexander \cite{alexander}; 
$^{\rm e}$ Miller \& Goodrich \cite{miller};
$^{\rm f}$ Tran et al. \cite{tran92};
$^{\rm g}$ Antonucci \& Miller \cite{antonucci};
$^{\rm h}$ V\'eron-Cetty \& V\'eron \cite{veronsy};
$^{\rm i}$ Kay \cite{kay};
$^{\rm j}$ Hines \& Wills \cite{hines};
$^{\rm k}$ Inglis et al. \cite{inglis};
$^{\rm l}$ Heisler et al. \cite{heisler};
$^{\rm m}$ Nagao et al. \cite{nagao};
$^{\rm n}$ Alexander et al. \cite{alexander99};
$^{\rm o}$ Moran et al. \cite{moran}. 
\end{table*}

\section{The Results}

\subsection{\nh versus detectability of polarized broad lines}

We show the histogram distribution of column densities for 
Sy2s without and with PBL in Figs. 1a and 1b,
respectively. Since there are 6 censored data 
(lower limits) among
Sy2s with PBL and 11 among Sy2s without PBL, we need to use 
the survival analysis methods (ASURV Rev 1.2, Isobe, Feigelson
\& Nelson \cite{isobe})
to study the similarity of these two samples.
We find that the probability 
for these two samples to be extracted from the same parent population
is about 12 \%, and the mean values of log~\nh (in units of cm$^{-2}$)
are 23.6 $\pm$ 0.2 and 24.2 $\pm$ 0.2,
respectively. This result suggests that Sy2s with PBL 
are affected by lower obscuration than Sy2s without PBL, but the
statistical significance of the result is not high, and therefore
not conclusive.

\subsection{Infrared colors versus \nh }

In Figure 2, we show the plot of
column density (\nha) versus
far-IR color ($\rm s_{60\mu m}/s_{25\mu m}$) for Seyfert 2 galaxies with 
PBL (filled circles) and Sy2s without PBL (open circles).
Sy2s with PBL have warmer FIR colors than Sy2s without PBL, which confirms the
result obtained by Heisler et al. (\cite{heisler}) with a higher statistical
significance, the probability for these two samples from the same 
parent population is less than 0.0001 \% and the mean values of $\log 
(\rm s_{60\mu m}/s_{25\mu m})$ are 0.368 $\pm$ 0.044 and  0.722 $\pm$ 0.053,
respectively.
However, we do not find any correlation between the
$\rm s_{60\mu m}/s_{25\mu m}$ color and the absorbing \nha, at variance
with what is expected from the model proposed by Heisler et al.
(\cite{heisler}): according to the latter model higher \nh should
correspond to cooler IR colors.

\begin{figure}
\vspace{0cm}
\epsfxsize=9.76cm
\centerline{\epsfbox{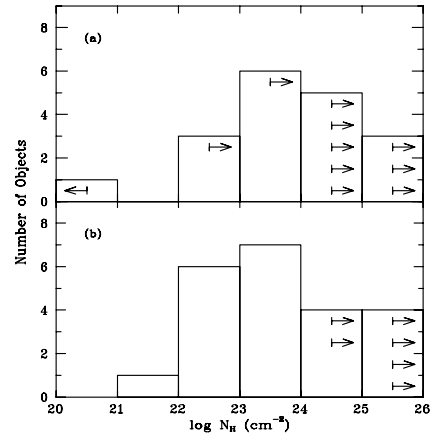}} 
\vspace{0cm}
\caption{Distribution of the absorbing column densities (\nha)
 for Seyfert 2 galaxies without polarized broad lines (Fig 1a)
  and for Sy2s with detected polarized broad lines
  (Fig 1b).}
\end{figure}

\begin{figure}
\vspace{0cm}
\epsfxsize=9.76cm
\centerline{\epsfbox{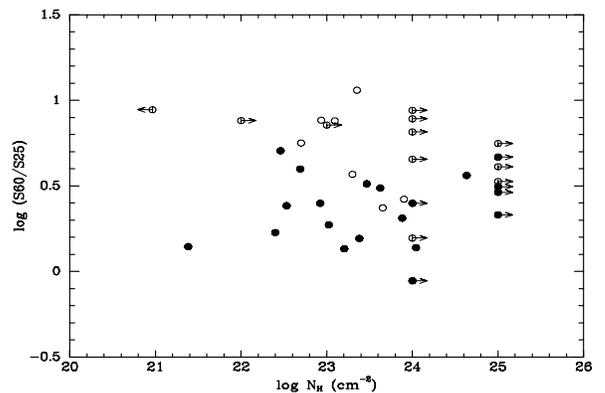}} 
\vspace{0cm}
\caption{Distribution of the absorbing column density (\nha) 
 versus the IR color ($\rm s_{60\mu m}/s_{25\mu m}$). Seyfert 2 galaxies 
 with and without PBL are
  marked with filled and open circles, respectively.}
\end{figure}

\begin{figure}
\vspace{0cm}
\epsfxsize=9.76cm
\centerline{\epsfbox{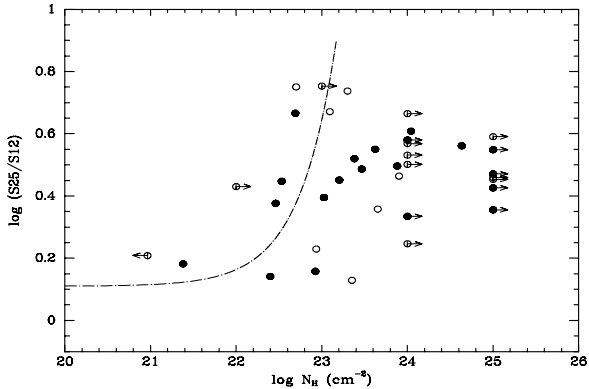}} 
\vspace{0cm}
\caption{Distribution of the absorbing column density (\nha) versus the IR
  color ($\rm s_{25\mu m}/s_{12\mu m}$). 
 Symbols have the same coding as in Fig.~2. The dot-dashed line indicates
the expected relation between IR color and absorbing \nh assuming that
the intrinsic IR color is the same as observed in Sy1s and that the dust
associated with the observed \nh completely cover the IR emitting region.}
\end{figure}

Fig.~3 is a plot similar to Fig.~2, but where the IR color is sampled
at shorter wavelengths and, more specifically, the
ratio $\rm s_{60\mu m}/s_{25\mu m}$ is replaced by
$\rm s_{25\mu m}/s_{12\mu m}$. This ratio should be more sensitive to
absorption of the inner (hotter) dust component.
In this case there is a marginal evidence
that Sy2s with \nh $ < 10^{23}$ cm$^{-2}$ have warmer colors. Yet, the relation
between $\rm s_{25\mu m}/s_{12\mu m}$ color and \nh is not the one
expected from the absorption by a dusty screen associated to the gaseous
column observed in the X-rays, assuming a Galactic gas-to-dust ratio and
extinction curve. Such a relation is given by the following formula:

\begin{equation}
\log (\frac{s_{25}}{s_{12}})_{obs} = \log (\frac{s_{25}}{s_{12}})_{int} + 0.042 \times N_{\rm H} \times 10^{-22}
\end{equation}

\noindent
where $\rm (s_{25\mu m}/s_{12\mu m})_{int}$ is the intrinsic color prior to
absorption and where we assumed

\begin{equation}
\rm 
A_{12} = \frac{1}{27} A_{\rm V} ~~~ and ~~~A_{25} = \frac{1}{55} A_{\rm V}
\end{equation}

\noindent
and 
\begin{equation}
A_{\rm v} = 5.0 \times 10^{-22} N_{\rm H}
\end{equation}

\noindent
(Bohlin et al. \cite{bohlin}).
To estimate the intrinsic $\rm s_{25\mu m}/s_{12\mu m}$ color, we
selected all of the Seyfert 1 galaxies listed in the V\'eron-Cetty \& V\'eron
(\cite{veronsy}) catalog with available IRAS fluxes and we derived 
a mean value of $\langle \log (\frac{s_{25}}{s_{12}}) \rangle = 0.11$.
Therefore, within the framework of the unified model we assigned this
value to $\log (\frac{s_{25}}{s_{12}})_{int}$ in equation 1. The
resulting curve is shown in Fig.~3 with a dot-dashed line, which may match the
observational data for low \nh ($< 10^{23}$ cm$^{-2}$),
but fails to account for the majority of the sources at higher columns.
This finding indicates that the observed
distribution of IR colors is not to be ascribed
to different degrees of absorption (at least for \nh $ > 10^{23}$ cm$^{-2}$).

\subsection{Detectability of polarized broad lines versus nuclear morphology}

HST images of a large sample of Seyfert galaxies have shown that Seyfert 2
galaxies are characterized by dust lanes or irregular dust
distribution crossing the nuclear region more often than Sy1s. Based on
these results, Malkan et al. (\cite{malkan}) suggested that 100pc-scale
dusty structures may play a role in the obscuration that generally affects
Sy2s. As summarized in Maiolino (\cite{maiolino}) the presence of a
 100pc-scale obscuring medium in Sy2s is supported by various pieces of
evidence, but most likely such a large scale medium
contributes to the absorption only with ``moderate''
gas columns (less than a few times 10$^{23}$ cm$^{-2}$).

Among all of the Seyfert 2 galaxies imaged by Malkan et al. (\cite{malkan})
we searched those observed in spectropolarimetry and found 32 of them.
Out of these 32 Sy2s, 12 have PBL and 20 do not have PBL. 30\% of the Sy2s
without PBL are characterized by (large scale) dust lanes crossing the
nuclear region or irregular nuclear dust distribution (according
to the classification given in Malkan et al. \cite{malkan}) while none of 
the Sy2s with
PBL show evidence for dusty nuclear features. This suggests that, at
least in some cases, obscuration due to 100pc-scale dusty structures
is responsible for hiding the mirror which reflects the broad lines.

\subsection{Detectability of the polarized broad lines on the Radio-FIR plane}

It is well known that a tight correlation exists between radio and far-IR
(FIR) emission for normal, starburst and Seyfert galaxies (the latter with
larger scatter) (Helou et al. \cite{helou}). More recently, Ji et al. (\cite{ji}) have studied the
radio--FIR relation of LINERs and found that the AGN-- and starburst--supported
LINERs can be distinguished on this diagram, with the AGN--dominated ones being
scattered in the region with higher radio fluxes with respect to the
starburst/normal galaxies correlation. In Fig.~4 we show the radio vs.
FIR diagram for the Sy2s observed in spectropolarimetry. The Sy2s with
PBL are spread mostly above the standard starburst correlation (dashed line),
indicating the presence of an extra contribution
to the radio emission due to the AGN. Instead, Sy2s without PBL follow
more tightly the starburst correlation, suggesting that in these objects
the starburst component dominates both the FIR and the radio emission.

\section{Discussion and Implications}

Our finding that the absorbing column density (\nha) is marginally lower
in Sy2s with PBL than in Sy2s without PBL is tentatively in favor
of Heisler's et al. (\cite{heisler}) model, which associates the visibility
of the PBL with the amount of obscuration along the
line of sight, though the statistical significance of the result is not high.
Yet, the interpretation of the correlation between detectability of PBL
and IR colors given by Heisler et al. is not supported by our results.
They ascribe the colder IR colors observed in objects without PBL to the larger
obscuration affecting the mid-IR emitting region. The mismatch between
the expected and the observed IR colors in Figs.~2--3 indicates that the
IR colors are unrelated to the obscuration affecting the active nucleus.
There are various possible scenarios (not necessarily alternative)
to explain such a mismatch.
For a Galactic gas-to-dust ratio
and extinction curve, at 12$\mu$m the extinction implied by a column
larger than $10^{23}$ cm$^{-2}$ is higher than 2 mag, implying that in this
case the 12$\mu$m radiation is heavily suppressed and, therefore,
the observed radiation is probably dominated by the host galaxy (especially
in the large IRAS beam). It is worth noting that at \nh $ \le 10^{23}$ cm$^{-2}$
the 25$\mu$m/12$\mu$m color is lower (i.e. hotter) and follows the relation
expected by the reddened Sy1 curve (dot-dashed curve in Fig.~3).
Indeed, in this range of low \nh 
the absorbing medium is more transparent to the 12$\mu$m radiation and might
well dominate over the emission from the host galaxy. At longer wavelengths
(25$\mu$m and 60$\mu$m) the dust extinction is much reduced and
the emitting region is much more extended. In particular, at large
gaseous columns (\nh $> 10^{24}$ cm$^{-2}$)
the medium responsible for absorption must
be very compact ($<$ 10 pc), not to violate constraints on the gas
mass given by the dynamical mass (Risaliti et al. \cite{risaliti}, Maiolino \cite{maiolino});
as a consequence, in many cases (at least in the Compton thick sources)
the obscuring medium is smaller than the dusty emitting region responsible for
the 25$\mu$m and 60$\mu$m radiation (10-100 pc).
Therefore, the scatter in the 60$\mu$m/25$\mu$m color probably reflects
mostly variations in the relative contribution of the AGN (hotter)
 and starburst/galactic (colder)
component to the IR radiation, as suggested by Alexander (\cite{alexander}),
although some obscuration effect on the 25$\mu$m emission might be present.

These findings confirm and strengthen the result obtained by Alexander
(\cite{alexander}) that the relation between visibility of PBL and IR colors is mostly due to the relative dominance of AGN and starburst/galactic
component; in the sense that the
latter both makes the IR colors cooler and dilutes the optical light making
more difficult the detection of the PBL. This scenario is further
supported by the finding that Sy2s without PBL follow the same
radio-FIR correlation as starbursts (sect.3.4).

Yet, we find that, although the
relation between PBL detectability and IR colors is mostly related
to the relative contribution of the starburst/galactic component,
the detectability of the PBL is also affected, to a lower extent, by the
obscuration toward the nuclear region. This is indicated by the larger
average \nh 
and by the higher incidence of nuclear dusty features in
Sy2s without PBL. Given the limited statistical significance of these
findings ($\sim$
90\% for the difference in \nh distribution) it is not surprising
that such a trend was not found by Alexander (\cite{alexander}) which used a
much smaller sample of objects.

Within the context of the relation between detectability of PBL and
dominance of the starburst component, there is a possible
explanation, alternative to the dilution of the optical light by the
host galaxy suggested by Alexander (\cite{alexander}), and which
might apply to some of the objects without PBL.
Some authors have suggested the existence of a dichotomy in Sy2s
where, at variance with the commonly accepted unified scenario,
a fraction of the Sy2s do not host a hidden Sy1 nucleus but are
intrinsically different, such objects would be ``pure'' Seyfert 2
galaxies (Hutchings \& Neff \cite{hutchings};
Neff \& Hutchings \cite{neff}; Heckman et al. \cite{heckman95}; 
Dultzin$-$Hacyan et al. \cite{deborah}; Gu et al. \cite{gu}). 
In particular the finding that the bolometric
luminosity of some objects with a Sy2 spectrum
is dominated by a nuclear starburst prompted Heckman et al. 
(\cite{heckman95}, \cite{heckman97}) to argue
that some extreme starburst population might mimic the narrow line
spectrum of AGNs (partly supporting the model of Terlevich et al. 
\cite{terlevich}).
Within this scenario, the relation
between detectability of PBL and IR colors is trivial: some of the Sy2s with
starburst-like cooler IR colors do not show evidence for PBL because the
broad line region is absent.

\begin{figure}
\vspace{0cm}
\epsfxsize=9.76cm
\centerline{\epsfbox{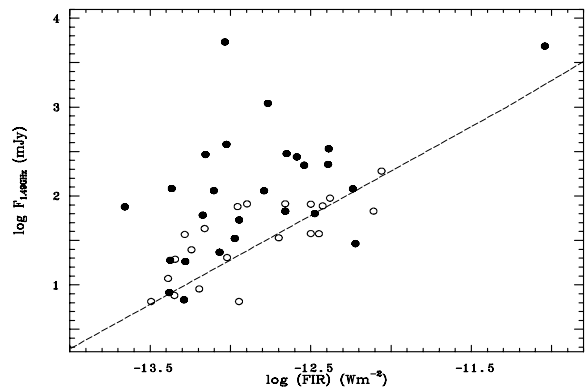}} 
\vspace{0cm}
\caption{Distribution of FIR and radio fluxes for the Sy2s observed
in spectropolarimetry.
Symbols have the same meaning as in Fig. 2. and the
 dashed line is the fit for normal and starburst
 galaxies from Helou et al. (\cite{helou}).} 
\end{figure}

\section{Conclusions}

In this paper, we collect 40 Seyfert 2 galaxies having both spectropolarimetric 
observations and a measure of the absorbing \nh obtained
by means of their X-ray spectrum. Out of these 40 objects 22 show
broad lines in polarization (most likely ascribed to scattering of the broad
line region) and 18 do not. For these objects we also analyzed the relation of
\nh and detectability of the broad lines with the mid- and far-IR colors.
We confirm previous claims that Sy2s without polarized broad lines have cooler
IR colors. We studied additional diagrams to test two scenarios
proposed to explain such a correlation and that, more specifically, ascribe
this effect either to obscuration of the nuclear region or to the
contribution/dilution from circumnuclear starburst activity and from the
host galaxy.

We found a marginal evidence for Sy2s showing broad lines in polarization to
have lower \nh. We also find that about 30\% of the Sy2s without polarized
broad lines show dust lanes crossing the nucleus, while none of the Sy2s
with polarized broad lines show evidence for such nuclear dusty structures.
These results suggest that the
obscuration towards the nuclear regions (hence towards
the scattering mirror) plays a role in hiding the polarized broad lines, at
least in some objects.

On the other hand, we find that the absorbing column density does not
correlate with the IR colors and, in particular, these quantities do not follow
the relation expected in the case of absorption of the IR emitting region by
the dust associated to the observed \nha, therefore 
indicating that the distribution of IR colors is not to be ascribed to
obscuration effects. Also, we find that Sy2s without detection of polarized
broad lines follow the same radio--FIR relation as normal and starburst
galaxies, at variance with Sy2s with polarized broad lines which tend to spread
towards higher radio luminosities. These findings support
previous claims that the relation between far-IR colors and
detectability of the polarized broad lines in Sy2s is mostly related to
dilution of the IR and optical light by a circumnuclear starburst  or by
the host galaxy.

Summarizing, our results indicate that the lack of broad lines in the
polarized spectrum of Sy2s is mostly due to the contribution/dilution
from the stellar component, though at a lower extent the
obscuration towards the nuclear region also
plays a role.

\begin{acknowledgements}
 We would like to thank the anonymous referee for his/her 
 careful reading the manuscript and valuable comments, which improved 
 the paper a lot.
 A significant fraction of this work was done during the Gullermo Haro
 Workshop 2000, we are grateful to the organizers of the workshop who
 made possible this collaboration and enabled us to perform this research.
 QSGU acknowledges support from UNAM postdoctoral program (Mexico) and
 from National Natural Science Fundation of China and the National Major 
 Project for Basic Research of the State Scientific Commission of China. 
 RM acknowledges partial support by the Italian Space Agency
 (ASI) under grant ARS--99--15 and by the Italian Ministry for
 University and Research (MURST) under grant Cofin98--02--32.
 And DD-H acknowledges support from grant IN 115599 from PAPIIT-UNAM. 
 This research
 has made use of NASA's Astrophysics Data System Abstract Service and
 the NASA/IPAC Extragalactic Database (NED) which is operated by the Jet
 Propulsion Laboratory, California Institute of Technology, under contract
 with the National Aeronautics and Space Administration.
\end{acknowledgements}

\end{document}